\documentstyle[dina4p,11pt,doublespace]{article}

\def\fn{\footnote}
\def\C{{\bf C}}
\def\bi{\bibitem}
\def\ti{\tilde}
\def\tr{{\rm tr}}
\def\i{\infty}
\def\E{{\cal E}}
\def\P{\Psi}
\def\x{\xi}
\def\xb{{\bar{\xi}}}
\def\vs{\vskip}
\def\be{\begin{equation}}
\def\ee{\end{equation}}
\def\la{\label}
\def\c{\cite}
\def\f{\frac}
\def\Eb{\bar{\cal E}}
\def\ba{\begin{array}}
\def\ea{\end{array}}
\def\r{\ref}
\def\l{w}
\def\O{\Omega}
\def\t{\gamma}
\def\omega{q}

\begin{document}

\newtheorem{Theorem}{Theorem}

\begin{center}

\LARGE{Separation of Variables and Hamiltonian Formulation
for the Ernst Equation}

\vskip2.0cm

\large{D.~Korotkin\fn{Supported by Alexander von Humboldt Foundation}
and H.~Nicolai}
\vskip1.0cm
II Institute for Theoretical Physics, Hamburg University \\
Luruper Chaussee 149, Hamburg 22761 Germany
\end{center}

\vs2.0cm
{\bf ABSTRACT.} It is shown that the vacuum Einstein equations for
an arbitrary stationary axisymmetric space-time can be completely
separated by re-formulating the Ernst equation and its
associated linear system in terms of a
non-autonomous Schlesinger-type dynamical system. The conformal factor
of the metric coincides (up to some explicitly computable factor)
with the
$\tau$-function of the Ernst equation in the presence of finitely
many regular singularities.
We also present a canonical formulation of these results, which
is based on a ``two-time" Hamiltonian approach, and which opens
new avenues for the quantization of such systems.
\vs2.0cm

{\bf Introduction.}
In this letter we demonstrate that the vacuum Einstein equations for
space-times with two commuting Killing vectors can be re-formulated
in terms of a pair of compatible {\it ordinary} matrix differential
equations. Similar results can be
shown to hold for the more general equations obtained by dimensional
reduction from higher-dimensional theories of gravity and supergravity
with matter couplings to two dimensions.
As a by-product, we establish a previously unknown
relation between the conformal factor of the associated metric
and the so-called $\tau$-function, which plays a pivotal role
in the modern formulation of integrable systems \c{J,New}. Thirdly, we
present a canonical formulation of these results, which avoids certain
technical difficulties encountered in previous treatments.
Our results suggest that an {\it exact} quantization of
axisymmetric stationary (matter-coupled) gravity by exploiting
techniques developed for flat space (quantum) integrable
systems \c{Fad,Korepin} is now within reach.

{\bf The Ernst equation and related linear system.}
We start from the following metric on stationary axisymmetric
space-time \c{Exact}
\be
ds^2=f^{-1}[e^{2k}(dx^2 +d\rho^2) +\rho^2 d\phi^2]-
f(dt + Fd\phi)^2
\la{m}\ee
where $(x,\rho)$ are Weyl canonical coordinates; $t$ and $\phi$ are
the time and angular coordinates, respectively. The functions
$f(x,\rho )$, $F(x,\rho)$ and $k(x,\rho)$ are related to the
complex Ernst potential $\E (x,\rho)$ by
\be
f={\rm Re}\E \;\;\; , \;\;\;
F_{\xi}=2\rho\f{(\E -\Eb)_{\xi}}{(\E+\Eb)^2}\;\;\; , \;\;\;
k_{\xi}= 2 i \rho\f{\E_{\xi}\Eb_{\xi}}{(\E+\Eb)^2}   ,
\la{mc}\ee
where $\x=x+i\rho\, ,\, \xb=x-i\rho$; hereafter subscripts $\x$, $\xb$
denote partial derivatives with respect to these variables.
In terms of the potential $\E(\x,\xb)$, Einstein's equations for
the metric (\r{m}) in particular imply the Ernst equation \c{Exact}
\be
((\x-\xb)g_{\x}g^{-1})_{\xb} + ((\x-\xb)g_{\xb}g^{-1})_{\x} =0
\la{ee}\ee
with the symmetric matrix
\be
g=\f{1}{\E+\bar{\E}}\left(\ba{rrr} 2 \;\;\;\;\;\;\; i(\E-\Eb)  \\  \\
                 i(\E-\Eb)\;\;\;\;\;\;\;\;  2\E\Eb \ea \right)  .
\la{g}\ee

The equation (\r{mc}) for $k(\x ,\xb )$ may be
equivalently written in the form
\be   d\log h =\omega   \la{f} \ee
where $h:= e^{2k}$ is the conformal factor. Using (\r{ee}) one can
show that the one-form $\omega$ defined by
\be
\omega=\f{\x-\xb}{4}\tr (g_{\x} g^{-1})^2 d\x +
\f{\xb-\x}{4}\tr (g_{\xb} g^{-1})^2 d\xb
\la{om}\ee
is closed, i.e. $d\omega =0$.
Equation (\r{ee}) is the compatibility condition of the following
linear system  \c{M,BZ}:
\be
\f{d\P}{d\x}=U\P  \;\;\;\;\;\;\;\;\;
\f{d\P}{d\xb}=V\P
\la{ls}\ee
where
\be
U=\f{g_{\x} g^{-1}}{1-\t}\;\;\;\;\;\;\;\;
V=\f{g_{\xb} g^{-1}}{1+\t}\;\; ;
\la{UV}\ee
and $\P(\t,\x,\xb)$ is a $2\times 2$ matrix, from which the Ernst
potential and thus the metric (\r{m}) can be reconstructed.
The function $\t(\x,\xb)$ is a ``variable spectral parameter"
subject to the following (compatible) first order equations
\be
\t_{\x}=\f{\t}{\x-\xb}\f{1+\t}{1-\t} \;\;\;\;\;\;\;\;
\t_{\xb}=\f{\t}{\xb-\x}\f{1-\t}{1+\t} .
\la{pe}\ee
They are solved by
\be
\t(\x,\xb,\l)
  =\f{2}{\x-\xb}\left\{\l-\f{\x+\xb}{2}-\sqrt{(\l-\x)(\l-\xb)}\right\}
\la{t}\ee
with $\l\in {\bf C}$ a constant of integration, which can be
regarded as the ``hidden" constant spectral parameter. For the
linear system (\r{ls}), we can use either $\t$ or $\l$; when
$\t$ is expressed as a function of $\l$ according to (\r{t}), the
linear system (\r{ls}) lives on the two-sheeted Riemann surface
of the function $\sqrt{(\l-\x)(\l-\xb)}$.
Both the constant and the variable spectral parameters $\l$
and $\t$ are needed for a proper understanding
of the (infinite-dimensional) hidden symmetries of (\r{ls}) \c{BM,Nic}.
Furthermore,
\be
\f{d}{d\x}\equiv\f{\partial}{\partial \x}+
\f{\t}{\x-\xb}\f{1+\t}{1-\t} \f{\partial}{\partial \t}\;\;\;\;\;\;\;\;\;
\f{d}{d\xb}\equiv\f{\partial}{\partial \xb}+
\f{\t}{\xb-\x}\f{1-\t}{1+\t} \f{\partial}{\partial \t}.
\la{cd}\ee
The poles of (\r{ls}) in the complex $\t$-plane are thus produced
by differentiation of $\t$ according to (\r{pe}).

Choosing $(\t,\x,\xb)$ as independent variables, we get
the following relations from (\r{ls}) and (\r{pe}) \c{NK}:
\be
g_{\x}g^{-1}=\f{2}{\x-\xb}\P_{\t}\P^{-1}|_{\t=1} \;\;\;\;\;\;\;\;
g_{\xb}g^{-1}=\f{2}{\x-\xb}\P_{\t}\P^{-1}|_{\t=-1}
\la{nk}\ee
where the subscript $\t$ denotes differentiation with respect to $\t$.

{\bf Deformation equations.}
Although we shall keep in mind the Ernst equation and its associated
linear system (\r{ls}) as our principal example, the results to
described below hold for arbitrary $GL(n,{\bf C})$-valued
matrices $g(\x,\xb )$, as well as for the gravitationally coupled
non-linear $\sigma$-models obtained by dimensional reduction
of Maxwell-Einstein theories in higher dimensions. For our
analysis, we shall use the general framework
of monodromy preserving deformations of ordinary differential
equations \c{J}.

Let us now consider the behavior of
$(d\P/d\x)\P^{-1}$ and $(d\P/d\xb)\P^{-1}$ in the complex $\t$-plane.
Singularities in $\t$ arise at those points where $\P (\t)$ is either
non-holomorphic or degenerate (i.e. $\det\P =0$).
Analyticity away from the points $\t =\pm 1$
implies that all singular points $\t_j$ (for $j=1,...,N$)
of the function $\P(\t,\x,\xb)$ are regular in the sense that \c{J}
\be
\P(\t)=G_j(\x,\xb)\P_j(\t,\x,\xb)(\t-\t_j)^{T_j} C_j\;\;\;\;{\rm as}\;\;
\;\;\t\sim \t_j
\la{rs}\ee
For $\t \sim \t_j$, $\P_j(\t,\x,\xb)={\bf 1} + O(\t-\t_j)$ is
holomorphic and invertible. The matrices $C_j$ and
$T_j$ are constant and invertible, and constant diagonal,
respectively, while the $(\x,\xb)$-dependent matrices $G_j$
are assumed to be invertible. The singular points $\t_j$
depend on $(\x,\xb)$ according to (\r{pe}), i.e. we have
$\t_j = \t(\l_j,\x,\xb)$ with constants $\l_j \in {\bf C}$ \c{1}.
The set $\{\t_j, C_j, T_j \}$ for $j=1,...,N$ is generally
referred to as the set of monodromy data
of $\P(\t)$. The function $\P(\t)$ is uniquely defined by its
monodromy data up to normalization \c{J}.

The logarithmic derivative $\P_{\t}\P^{-1}$ is thus holomorphic
except at the points $\t=\t_j$ where it has simple poles with residues
\be
A_j (\x,\xb) =G_j T_j G_j^{-1}
\la{Aj}\ee
by (\r{rs}). The functions $A_j(\x,\xb)$ will play a central role
in the sequel. In general the number $N$ of regular singularities
$\t_j$ may be infinite (explicit examples are the $x$-periodic
static axisymmetric solutions found in \c{Myers}) or even continuous
(this would correspond to non-constant conjugation matrices in the
related Riemann-Hilbert problem).
However, in this paper we will restrict attention to
finite $N$. Besides that, we find it convenient to impose
the normalization condition
$\P_{\t}\P^{-1}|_{\t=\i}= 0$, which may be ensured for instance by
demanding $\P |_{\t=\i}= \sigma_3 $. A large class of solutions with
finitely many singularities is provided by the multisoliton
solutions of Einstein's equations in \c{BZ} (corresponding to matrices
$T_j$ all of whose eigenvalues are half-integer) and the finite-gap
(algebro-geometric) solutions constructed in \c{KM}.

Combining (\r{rs}) and (\r{Aj}) we arrive at the following
differential equation in $\t$:
\be
\frac{\partial \P}{\partial \t} =\sum_{j=1}^{N}\f{A_j}{\t-\t_j}\P
\la{de}\ee
Inserting (\r{de}) into (\r{nk}), we immediately obtain
\be
g_{\x} g^{-1}=\f{2}{\x-\xb}\sum_{j}\f{A_j}{1-\t_j}\;\;\;\;\;\;\;\;
g_{\xb} g^{-1}=\f{2}{\xb-\x}\sum_{j}\f{A_j}{1+\t_j}
\la{cur}\ee
(in the sequel summation is taken everywhere from $1$ to $N$).
Substituting (\r{cur}) into (\r{ls}), we get the following
compatibility conditions between (\r{ls}) and (\r{de}) \c{Sch}:
\be
\f{\partial A_j}{\partial\x}=
        \f{2}{\x-\xb}\sum_{k\neq j}\f{[A_k,\;A_j]}{(1-\t_k)(1-\t_j)}
\;\;\;\; , \;\;\;\;
\f{\partial A_j}{\partial\xb}=
\f{2}{\xb-\x}\sum_{k\neq j}\f{[A_k,\;A_j]}{(1+\t_k)(1+\t_j)}
\;\;\;\;\;\;\;\;      (j=1,...,N)
\la{1}\ee
It is now straightforward to check that this system is always
compatible if the functions $\t_j$ obey (\r{pe}).
The first main result of this letter is thus

\begin{Theorem}
Let $\{\l_j\in{\bf C}\; ; \;j=1,...,N\}$ be an arbitrary set of complex
constants and $A_j = A_j(\x,\xb)$ an associated set of solutions of
(\r{1}). Then the system of linear equations (\r{cur}) is always
compatible, and the matrix function $g=g(\x,\xb)$
obtained by integrating (\r{cur}) solves (\r{ee}).
\end{Theorem}
The proof may be obtained by direct calculation.
\medskip

It is quite remarkable that the dependence
of the Ernst equation and its associated linear system on the
variables $\x$ and $\xb$ can be completely decoupled by this theorem.
In other words, the problem of solving Einstein's equations in
this reduction has been reduced to integrating two {\it ordinary}
matrix differential equations, which are automatically
compatible unlike the original linear system (\r{ls})! All information
about the degrees of freedom is thereby encoded into the
``initial values", i.e. the set of (constant) matrices
$A_j^{(0)} \equiv A_j(\x^{(0)}, \xb^{(0)})$;
these are also the appropriate phase
space variables, as we will see below. Accordingly, we will
regard the functions $A_j(\x,\xb)$ rather than $\P(\t,\x,\xb)$ as
the fundamental quantities from now on, and relate the system
(\r{1}) directly to the (complexified) Ernst equation (\r{ee}).

Equations (\r{1}) may also be represented in ``Lax form", viz.
\be
\frac{\partial A_j}{\partial \x} =
   \Big[ U|_{\t=\t_j},A_j\Big]  \;\;\;\;\;\; ,  \;\;\;\;\;\;\;
\frac{\partial A_j}{\partial \xb} =
   \Big[ V|_{\t=\t_j},A_j\Big]  ,
\la{1a}\ee
where the matrices $U$ and $V$ are defined in (\r{UV}) and (\r{cur}).
This form of (\r{1}) is ``gauge-covariant" with respect
to the transformation
\be
\ti{\P}=\O(\x,\xb)\P,
\la{gau}\ee
Namely, the transformed function $\ti \P$ satisfies the linear system
$d\ti{\P}/d\x = \ti{U}\ti{\P} ,\;\;
d\ti{\P}/d\xb = \ti{V}\ti{\P},$
where
\be
\ti{U}=\O_{\x}\O^{-1}+\O U\O^{-1} \;\;\;\;\;\; , \;\;\;\;\;\;
\ti{V}=\O_{\xb}\O^{-1}+\O V\O ^{-1}
\la{UV1}\ee
Clearly, the matrix functions $A_j$ transform as
$A_j\rightarrow \ti{A}_j=\O A_j\O^{-1}$ under (\r{gau}).
The transformed matrices $\ti{A}_j$ then obey the same linear
system (\r{1a}) with the pair $(U,V)$ replaced by
$(\ti{U},\ti{V})$.

Theorem 1 establishes a direct correspondence between
$GL(2, {\bf C})$-valued
solutions of (\r{ee}) and solutions of (\r{1}). However, it does
not specify the conditions that must be imposed on $\{\l_j\, ,\, A_j \}$
in order satisfy certain restrictions which the metric $g$ may be
subject to. It is easy to see that the condition
$\det g=1$ is guaranteed by $\tr A_j =0$; reality of $g$ requires the
existence of an involution (complex conjugation) on the set
$\{\l_j,\;A_j\}$. Conditions that ensure $g=g^T$ are more difficult
to formulate and will be discussed elsewhere.

{\bf Conformal factor and $\tau$-function.}
To each solution $\{A_j\}$ of (\r{1})
we can associate the following closed one-form \c{J}:
\be
\omega_0(\x,\xb)=\sum_{j\neq k} \tr (A_j A_k)d\log
(\t_j-\t_k)
\la{om0}\ee
where the exterior derivative $d$ is to be taken with respect to the
deformation parameters $(\x,\xb)$. The closure condition
$d\omega_0 =0$ may be directly verified by use of (\r{1}) and (\r{pe}).
Following the general prescription given in \c{J}, we define
the $\tau$-function associated with the Ernst equation by
\be
d\log\tau =\omega_0
\la{tau}\ee

We will now show that this $\tau$-function has a very definite
physical meaning in our context: up to an explicit factor, it
is just the conformal factor $h\equiv e^{2k}$ \c{Kit}!
To establish this result, we first substitute
(\r{cur}) into (\r{om}); then using (\r{pe}) and (\r{om0}) we obtain
\be
 \omega=\omega_0+\f{1}{\x-\xb}\sum_{j}\tr A_j^2
\left\{\f{d\x}{(1-\t_j)^2} -\f{d\xb}{(1+\t_j)^2}\right\}+
\sum_{j<k}\tr (A_j A_k)d\log (\x-\xb)
\la{f1}\ee

Now, from (\r{1}) it follows that
$\sum_{j} A_j$ is $(\x,\xb)$-independent. Furthermore, it is easy
to check that $\tr A_j$ and $\tr A_j^2$ are independent of $\x$ and
$\xb$, hence constant, for all $j$. Therefore, the eigenvalues of
$A_j$ are $(\x,\xb)$-independent for any solution of (\r{1})
that agrees with (\r{Aj}). As a consequence, the expression
$ \sum_{j<k}\tr (A_j A_k)$
is likewise $(\x,\xb)$-independent, and all extra terms on the
r.h.s. of (\r{f1}) may be explicitly integrated.

Using (\r{pe}) and (\r{t}), we thus arrive at

\begin{Theorem}
The conformal factor $h$ (\r{f}) and the $\tau$-function (\r{tau})
are related by
\be
h(\x,\xb,\l_j) =
C(\x-\xb)^{\f{1}{2}\tr\left\{\sum_{j} A_j\right\}^2}\prod_j
\left\{\f{\partial \t_j}{\partial\l_j}\right\}^{\f{1}{2}\tr A_j^2}
\tau (\x,\xb,\l_j)
\la{link}\ee
where $C\in {\bf C}$ is a constant of integration.
\end{Theorem}
Notice once more that quantities $\tr(\sum_{j} A_j)^2$ and $\tr A_j^2$
are $(\x,\xb)$-independent. If the related matrix $g$ is real and
symmetric, then $\sum_j A_j =0$, and the first factor
on the r.h.s. of (\r{link}) drops out. We emphasize that our result
is more general than previous ones
(the explicit computability of $h$ for multi-soliton solutions
has been known for a long time \c{BZ}), and valid
for arbitrary non-linear $\sigma$-models coupled to gravity.

{\bf Hamiltonian formulation.}
The system (\r{1}) is a ``two-time"
hamiltonian system with respect to the standard
Lie-Poisson bracket \c{Fad, Harnad}
\be
 \left\{A(\t)  \stackrel{\otimes}{,}  A(\mu)\right\} =
\Big[ r(\mu-\t)\, ,\, A(\t)\otimes {\bf 1}+{\bf 1} \otimes A(\mu)\Big]
\la{pb}\ee
where $A(\t)\equiv \P_{\t}\P^{-1}$ and the classical rational
$R$-matrix $r(\t)$ is equal to $\Pi/\t$ with $\Pi$ the
permutation operator in $\C^2\times \C^2$. The dynamics in the $\x$
and $\xb$-directions are governed by the Hamiltonians
\be
H_1 \equiv 2k_\x = \f{1}{\x-\xb}\sum_{k,\;j}\f{\tr(A_j A_k)}
{(1-\t_j)(1-\t_k)} \;\;\;\;\;\; , \;\;\;\;\;\;
H_2 \equiv 2k_\xb = \f{1}{\xb-\x}\sum_{k,\;j}\f{\tr(A_j A_k)}
{(1+\t_j)(1+\t_k)}      \la{Ham}
\ee
Compatibility of the system (\r{1}) implies
$\{H_1 , H_2\}=0$, i.e. the flows with respect to the
two ``time variables" $\x$ and $\xb$ commute (as can also
be verified by explicit computation). Note that our formulation
is far simpler both technically and conceptually than previous
hamiltonian treatments of such systems, which were based on the
use of ``one-time" Hamiltonians, and where the Lie-Poisson brackets
(\r{pb}) would in addition depend on the space coordinates.
The notorious problems caused by derivatives
of $\delta$-functions (so-called ``non-ultralocal" terms) in the
relevant Poisson brackets are altogether avoided here. Furthermore,
the cumbersome structure of the canonical current algebra
in the conventional approach is replaced by a more transparent
algebraic structure in our framework.

To be sure, we should regard (\r{Ham}) as constraints \`a la Dirac
rather than conventional Hamiltonians, because (\r{ee}) is derived
from a {\it generally covariant} theory. To do this properly
would, however, require that we undo the choice of Weyl coordinates,
on which (\r{m}) and (\r{ee}) are based, and to treat $\x$ and $\xb$
as canonical variables subject to
$\{ 2k_\x , \x \} = \{ 2k_\xb , \xb \} =1$
and $\{ 2k_\x , \xb \} = \{ 2k_\xb , \x \} =0$.
The quantities $\Phi_1 := 2k_\x - H_1$ and
$\Phi_2 := 2k_\xb - H_2$ are thereby converted into (mutually
commuting) constraint operators on an enlarged phase space
(they are, in fact, just the Virasoro constraints). The ``time
evolutions" of the spectral parameter $\t$ are also generated
canonically in the sense that $\t_\x = \{ \Phi_1 , \t \}$
and $\t_\xb = \{\Phi_2 , \t\}$
(so ``time" must eventually be quantized in this scheme!).

{\bf Final remarks.}

{\bf 1.} An obvious advantage of using the variables
$A_j$ in comparison with the ones employed traditionally in this
context is that they generate a closed Lie algebra
with respect to the standard Lie-Poisson bracket (\r{pb}).
Secondly, the common features of our system (\r{1}) with
the classical limit of the Knizhnik-Zamolodchikov equations
\c{Harnad,Reshet} suggest that one should quantize (\r{1}) in analogy
with the KZ equations (although quantum gravity will certainly
introduce new features). We have only discussed the case of finite
$N$ in this paper, but there are in principle no obstacles
to considering $N=\i$ from the outset, since the space of finite
$N$ solutions can be naturally embedded in this larger space.
For illustrative purposes, it is quite useful to think of the set of
finite $N$ solutions as the ``$N$-particle sector" of the theory
because, depending on the reality conditions, the solutions $g$
corresponding to (\r{de}) possess exactly $N$ or $\f{N}{2}$
singularities $\l_1,...,\l_N$ in the upper $\x$-half-plane.
In fact, a proper treatment of the Ernst equation as a (generally
covariant) quantum field theory will presumably require taking
into account $N$ as a ``particle number operator".

{\bf 2.} The extension of our results to the case of a Lorentzian
world-sheet and to arbitrary $G/H$ coset space $\sigma$-models
coupled to gravity in two space-time dimensions is straightforward.
In the notation of \c{Nic}, where this case is reviewed in some
detail, the linear system matrix $\P$ corresponds to
$\widehat {{\cal V}} \eta ({\cal V})^{-1}$, where
$\eta$ denotes the Cartan involution (e.g. $\eta({\cal V}) =
({\cal V}^T)^{-1}$ for $G=SL(n)$); furthermore, the spectral
parameter $t$ used there corresponds to $i$ times the parameter $\t$
employed in the present paper.
For arbitrary coset spaces, the matrices $A_j$ belong to the
Lie-algebra of $G$; like $g$, they may be subject to further
restrictions.

{\bf 3.} Analogs of the static axisymmetric (multi-Schwarzschild)
solutions for arbitrary $\sigma$-models can be easily constructed in
our formalism by choosing the matrices $A_j$ in the
Cartan subalgebra of the relevant Lie algebra. From (\r{1}) it is
then immediately evident that $A_j =const$.

{\bf 4.} Obviously our formulation will yield a new
realization of the Geroch group \c{Ger} and its generalizations;
we here just note that this group mixes sectors belonging to
different ``particle numbers". It is known that the
corresponding Kac Moody algebras act on the conformal factor via
their central extension \c{Jul, BM, Nic}; combining this result with
our Theorem 2 should shed some light on the group theoretical
meaning of the $\tau$-function.
\medskip

\noindent
A detailed account of the results described in this letter is
in preparation.
\medskip

\noindent
{\bf Acknowledgements.}
We thank Alexander Kitaev for important discussions.

\end{document}